\documentclass[12pt]{amsart}
\usepackage{graphicx,amsmath,amssymb}

\newcommand{\C}{\mathbb C}
\newcommand{\R}{\mathbb R}
\newcommand{\Z}{\mathbb Z}
\newcommand{\N}{\mathbb N}
\renewcommand{\d}{\prime} 
\newcommand{\dd}{{\prime \prime}}

\renewcommand{\Re}{{\rm Re}\,}
\renewcommand{\Im}{{\rm Im}\,}
\newtheorem{theorem}{Theorem}
\newtheorem{lemma}[theorem]{Lemma}
\newtheorem{proposition}[theorem]{Proposition}

\newtheorem{definition}{Definition}
\newtheorem*{remark}{Remark}

\begin{document}
\title[]
{The potential $(iz)^m$ generates real eigenvalues only, under symmetric rapid decay conditions}
\author[]
{K. C. Shin}
\address{Department of Mathematics, University of Illinois, Urbana, IL 61801}
\date{July 18, 2002}

\begin{abstract}
We consider the eigenvalue problems $-u^\dd(z)\pm (iz)^m u(z)=\lambda u(z)$, $m\geq 3$, under every rapid decay boundary condition that is symmetric with respect to the imaginary axis in the complex $z$-plane. We prove that the eigenvalues $\lambda$ are all positive real. 
\end{abstract}

\maketitle

\begin{center}
{\it Preprint.}
\end{center}

\baselineskip = 18pt

\section{Introduction}
\label{introduction}
For integers $m\geq 3$ fixed and  $1\leq \ell \leq m-1$, we are considering the non-Hermitian eigenvalue problems
\begin{equation}\label{H_peq}
H_{\ell}u(z):=\left[-\frac{d^2}{dz^2}+(-1)^{\ell} (iz)^m\right]u(z)=\lambda u(z),\quad\text{for some $\lambda\in\C$},
\end{equation}
with the boundary condition that 
\begin{equation}\label{bdcond}
\text{$u(z)\rightarrow 0$ exponentially, as $z\rightarrow \infty$ along the two rays}\quad \arg z=-\frac{\pi}{2}\pm \frac{\ell+1}{m+2}\pi.
\end{equation}
If a non-constant function $u$ along with a complex number $\lambda$ solves (\ref{H_peq}) with the boundary condition (\ref{bdcond}), then we call $u$ an {\it  eigenfunction} and $\lambda$  an {\it eigenvalue.} 
It is known that any solution of (\ref{H_peq}) is entire (analytic on the complex plane).
And it is also known that every solution of (\ref{H_peq}) is either decaying to zero or blowing up exponentially as $z$ tends to infinity along any ray $\{z\in \C: \arg z=const.\}$, except along $m+2$ critical rays where the transition between decaying and blowing-up sectors might occur \cite[\S 7.4]{Hille}. Along these $m+2$ critical rays, any non-constant solution is decaying algebraically \cite[\S 7.4]{Hille}. We will explain these asymptotic properties of the solution in Section \ref{properties}.

Before we state our main theorem, we first introduce some known facts  about the eigenvalues $\lambda$ of $H_{\ell}$, facts due to Sibuya \cite{Sibuya} and Hille \cite{Hille}.
\begin{proposition}\label{main2}
For $1\leq \ell\leq m-1$, the eigenvalues $\lambda$ of $H_{\ell}$ are the zeros of a certain entire function of order $\frac{1}{2}+\frac{1}{m}\in\left(\frac{1}{2},1\right)$. In particular, the eigenvalues have the following properties.
\begin{enumerate}
\item[(i)] Eigenvalues are discrete.
\item[(ii)] All eigenvalues are simple.
\item[(iii)] Infinitely many eigenvalues exist.
\end{enumerate} 
\end{proposition}
For our purposes, we will need to examine the proof of this proposition in some details. 
In Lemmas \ref{equiv} and \ref{equiv1}, we prove that the eigenvalues are zeros of certain entire functions of order $\frac{1}{2}+\frac{1}{m}\in\left(\frac{1}{2},1\right)$. Then the claims (i) and (iii) are consequences of the Hadamard factorization theorem, while the claim (ii) is due to Hille \cite[\S 7.4]{Hille}.

 In this paper, we will prove the following main theorem regarding the positivity of the eigenvalues.
\begin{theorem}\label{main1}
The eigenvalues $\lambda$ of $H_{\ell}$ for integers $1\leq \ell\leq m-1$ are all positive real.
\end{theorem}
The eigenvalues of $H_{\ell}$ are the same as those of $H_{m-\ell}$, as we show in the proof by reflecting $z\mapsto -z$.
The case $\ell=1$ of the theorem is due to  Dorey, Dunning and Tateo \cite{Dorey}, and we use this in our proof. Also when $m$ is even and  $\ell=\frac{m}{2}$, one can see that  $H_{\ell}$ is Hermitian on the real line, and hence $\lambda\in \R$.  In all other cases  Theorem \ref{main1} is new, and provides the first result of its kind for  boundary conditions that neither cluster near the negative imaginary axis nor lie on the real axis. 
We will explain how Theorem \ref{main1} covers {\it every} symmetric rapid decay condition later  when we discuss admissible boundary conditions in Section \ref{properties}.

For the rest of the Introduction, we will mention brief history and give some background about our main problem and our method of proof, and then in Section \ref{properties}, we will introduce work of Hille \cite[\S 7.4]{Hille} and Sibuya \cite{Sibuya} about some properties of solutions of (\ref{H_peq}). In Section \ref{induction_step}, we establish an induction step on $\ell$, which is the key element in our proof of Theorem \ref{main1}. More precisely, we will prove that for $1\leq \ell\leq\frac{m}{2}-1$, every eigenvalue $\lambda$ of $H_{\ell+1}$ is positive real if all eigenvalues $\sigma$ of $H_{\ell}$ lie in the sector $|\arg \sigma|\leq\frac{2\pi}{m+2}$ in the complex plane.  In Section \ref{main_proof}, we will outline a proof of the induction basis $\ell=1$, that is, eigenvalues of $H_1$ lie in the sector $|\arg \lambda|\leq\frac{2\pi}{m+2}$  (in fact $\arg \lambda=0$; see \cite{Dorey} and also \cite{Shin2}). We then prove Theorem \ref{main1} by induction on $\ell$ and the reflection $z\mapsto -z$. In Section \ref{discussion_de}, we discuss some hopes and challanges in extending our method to more general polynomial potentials. Finally, in the last section, we mention some open problems. 
 
\subsection*{History and overview of the method}
In this subsection, we introduce some earlier work related with our main result, Theorem \ref{main1}. Also, we discuss our method of the proof of Theorem \ref{main1}.

The Hamiltonians with the potential $\pm (iz)^m$ have been studied in many physics and mathematics papers, either under a boundary condition on the real axis \cite{CGM,Simon}, $u(\pm \infty+0i)=0$,  or under the boundary condition (\ref{bdcond}) with $\ell=1$ \cite{Bender,Dorey,Shin2}.

Simon \cite{Simon} and  Caliceti et al.\ \cite{CGM} studied the Hamiltonians  $-\frac{d^2}{dx^2}+x^2+\beta x^{m}$ on the real line, where $\beta\in \C-\R_-,$ $m=3,4,5,\dots$, and they proved compactness of the resolvent and discreteness of spectrum. Regarding the reality of eigenvalues,  Caliceti et al.\ \cite{CGM} showed that eigenvalues for $-\frac{d^2}{dx^2}+x^2+\beta x^{2n+1}$ are real if $\beta$ is small enough.

Recently, a conjecture of Bessis and Zinn-Justin  has been verified by Dorey et al.\ \cite{Dorey}, (and extended by the author \cite{Shin2}). That is, the eigenvalues $\lambda$ of 
\begin{equation}\label{zinn}
\left[-\frac{d^2}{dz^2}-\alpha(iz)^{3}+\beta z^2\right]u(z)=\lambda u(z),\quad u(\pm\infty+0i)=0\quad\text{for}\quad \alpha\in \R-\{0\},\,\,\beta\in \R,
\end{equation}
are all positive real. Dorey  et al.\ \cite{Dorey} verified for the case $\beta=0$, and later the author extended for the case $\beta\in\R$ \cite{Shin2}.

In fact,  Dorey et al.\ \cite{Dorey} have proved more. They studied the following eigenvalue problem
\begin{equation}\label{bessis}
\left[-\frac{d^2}{dz^2}-(iz)^{2M}-\alpha(iz)^{M-1}+\frac{l(l+1)}{z^2}\right]u(z)=\lambda u(z),
\end{equation}
 under the boundary condition (\ref{bdcond}) with $\ell=1$, and $M, \alpha, l$ being all real. They proved that for $M>1,$ $\alpha<M+1+|2l+1|$, the eigenvalues are all real, and for $M>1,$ $\alpha<M+1-|2l+1|$, they are all positive. A special case of (\ref{bessis}) is the potential $iz^3$ (when $M=\frac{3}{2},\,\alpha=l=0$), which is the $\beta=0$ version  of the Bessis and Zinn-Justin conjecture, but their results do not cover the $\beta\not=0$ version.

Later, the author \cite{Shin2} studied the following eigenvalue problem. 
\begin{equation}\nonumber
 -\left[\frac{d^2}{dz^2}+(iz)^m+a_1(iz)^{m-1}+a_2(iz)^{m-2}+\cdots+a_{m-1}(iz)\right]u(z)=\lambda u(z),
\end{equation}
with the boundary condition (\ref{bdcond}) with $\ell=1$, where $a_k\in\R$ for all $k$.
He proved that if for some $1\leq j\leq\frac{m}{2}$ we have $(j-k)a_k\geq 0$ for all $k$, then the eigenvalues $\lambda$  are all positive real \cite[Theorem 2]{Shin2}. This covers the full BZJ conjecture.

The proof of our main theorem, Theorem \ref{main1}, has four parts. The first part follows closely the method of Dorey et al.\ \cite{Dorey,Dorey2}, developing functional equations for spectral determinants, expressing them in factored forms and estimating eigenvalues by Green's transform.  The second part  establishes an induction step on $1\leq\ell\leq\frac{m}{2}$ and estimates eigenvalues by Green's transform again. In the third part, we use the result of  Dorey et al.\ \cite{Dorey2} that says the eigenvalues of $H_1$ are all positive real (see also Theorem 2 in \cite{Shin2}). This induction basis, along with the induction step established in the second part, proves our main theorem for $1\leq\ell\leq\frac{m}{2}$. Lastly we use the reflection $z\mapsto -z$ to cover $\frac{m}{2}<\ell\leq m-1$. Of course both this paper and \cite{Dorey,Dorey2,Shin2} are indebted to the work of Sibuya \cite{Sibuya}.

\subsection*{$\mathcal{PT}$-symmetric oscillators}
The above Hamiltonians are not  Hermitian in general, and hence the reality of eigenvalues is not obviously guaranteed. However, these  Hamiltonians share a common symmetry, the so-called $\mathcal {PT}$-symmetry.
A $\mathcal {PT}$-symmetric Hamiltonian is a  Hamiltonian which is invariant under the combination of the parity operation $\mathcal P(: z \mapsto -\overline{z})$ (an upper bar denotes the complex conjugate) 
and the time reversal operation $\mathcal T(: i \mapsto
-i)$.   These $\mathcal {PT}$-symmetric Hamiltonians have arisen in recent years in a number of physics papers, see \cite{Bender3,Handy2,Handy1,mez,Ali1,Znojil} and other references mentioned above, which support that some $\mathcal {PT}$-symmetric Hamiltonians have real eigenvalues only. The work of Dorey et al.\ \cite{Dorey} and the author \cite{Shin2}, and the results in this paper, prove rigorously that some $\mathcal {PT}$-symmetric Hamiltonians indeed have real eigenvalues only.

We also know that if $H=-\frac{d^2}{d z^2}+V(z)$ is
$\mathcal{PT}$-symmetric and $V(z)$ is a polynomial, then $V(z) =Q(iz)$ for some real polynomial $Q$, because $\overline{V(-\overline{z})}=V(z)$ and so $\Re V(z)$ is
an even function and $\Im V(z)$ is an odd function. Certainly (\ref{H_peq}) is a $\mathcal {PT}$-symmetric Hamiltonian. Moreover, if $u(z)$ is an eigenfunction of $H_{\ell}$ with the corresponding eigenvalue $\lambda\in\C$, then so is $\overline{u(-\overline{z})}$, with the corresponding eigenvalue $\overline{\lambda}$.

\subsection*{The exceptional symmetric boundary condition}
We will explain admissible boundary conditions in more detail in Section \ref{properties}. After that it will be clear that besides (\ref{bdcond}) the only other exponentially decaying boundary condition  that is symmetric with respect to the imaginary axis is that $m$ is even and $u(z)$ decays to zero as $z\rightarrow\infty$ along the both ends of the imaginary axis. 

But if $u(z)$ is an eigenfunction of $-u^\dd(z)-(iz)^mu(z)=\lambda u(z)$ satisfying this ``exceptional'' boundary condition, then
 $v(z):=u(-iz)$ is an eigenfunction of the Hermitian equation
$$-v^\dd(z)+z^{m}v(z)=-\lambda v(z),\quad v(\pm \infty+0i)=0.$$
So $\lambda\in\R$.

The existence of the eigenvalues in this case is clear from Proposition \ref{main2} applied to  the above equation of $v$.

\section{Properties of the solutions}
\label{properties}
In this section we will introduce some definitions and known facts related with the equation (\ref{H_peq}). One of our main tasks is to identify the eigenvalues as being zeros of certain entire functions, in Lemmas \ref{equiv} and \ref{equiv1}. But first, we rotate the equation (\ref{H_peq}) as follows because some known facts, which are related to our argument throughout, are directly available for this rotated equation.

Fix the integer $m\geq 3$.
Let $u$ be a solution of (\ref{H_peq}) and let $v(z)=u(-iz)$. Then $v$ solves
\begin{equation}\nonumber
-v^\dd(z) +\left[(-1)^{\ell+1} z^m+\lambda\right]v(z)=0.
\end{equation}
And the boundary condition (\ref{bdcond}) of $u$ becomes that $v(z)\rightarrow 0$ exponentially as $z\rightarrow \infty$ along the two rays $$\arg z=\pm \frac{\ell+1}{m+2}\pi.$$ 
Throughout this paper, we will use the complex number
\begin{equation}\nonumber
\fbox{$\omega=\exp\left[\frac{2\pi i}{m+2}\right].$}
\end{equation}
When $\ell$ is even,  it will be convenient to rotate once more, letting $w(z):=v(\omega^{\frac{1}{2}}z)$ so that $w(z)$ solves 
$$-w^\dd(z) +\left[z^m+\omega\lambda\right]w(z)=0.$$
Hille \cite[\S 7.4]{Hille} and Sibuya \cite{Sibuya} have studied the general equation of the form 
\begin{equation}\nonumber
-v^\dd(z) +\left[z^m+P(z)+\lambda\right]v(z)=0,
\end{equation}
where $P(z)$ is a polynomial of degree less than $m$.

We proceed now to summarize work of  Hille \cite[\S 7.4]{Hille}, and expand on some work of Sibuya \cite{Sibuya} for $P\equiv 0$, that is for
\begin{equation}\label{rotated}
-v^\dd(z) +\left[z^m+\lambda\right]v(z)=0,\quad\text{where}\quad m\geq 3.
\end{equation}

\subsection*{Results of Hille}
\begin{figure}[t]
    \begin{center}
    \includegraphics[width=.4\textwidth]{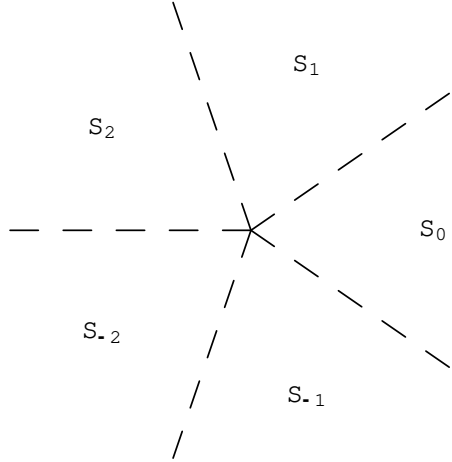}
    \end{center}
 \vspace{-.5cm}
\caption{The Stokes sectors $S_j$ of (\ref{rotated}) with $m=3$. The dashed rays are the critical rays: $\arg z=\pm\frac{\pi}{5},\,\pm\frac{3\pi}{5},\, \pi.$}\label{f:graph1}
\end{figure}

It is known that every solution of (\ref{rotated}) has simple asymptotic behavior near infinity \cite[\S 7.4]{Hille}. We will explain this asymptotic behavior using the following. 
\begin{definition}
{\rm Consider the equation 
\begin{equation}\label{rossi}
g^\dd(z)+\left[b_mz^m+b_{m-1}z^{m-1}+\cdots+b_1z+b_0\right]g(z)=0,
\end{equation} 
where  $b_k\in \C$ for $0\leq k\leq m$ with $b_m\not=0$. Let
$$\theta_j=\frac{2j\pi-\arg b_m}{m+2},\quad\text{for $j\in \Z$},$$
where we choose $-\pi<\arg b_m\leq \pi$.
For $j\in \Z$ we call the open sectors
\begin{equation}\label{gen_stokes}
S_j=\left\{z\in \C:\theta_j<\arg z<\theta_{j+1}\right\}
\end{equation}
the {\it Stokes sectors} of  (\ref{rossi}).
 Also we call the rays $\left\{\arg z=\theta_j\right\}$ the {\it critical rays.}

In particular, the Stokes sectors of (\ref{rotated}) are
\begin{equation}\label{stokes_sector}
S_j=\left\{z\in \C:\frac{(2j-1)\pi}{m+2}< \arg z<\frac{(2j+1)\pi}{m+2}\right\},\quad\text{for $j\in\Z$}.
\end{equation}
See Figure \ref{f:graph1}.}
\end{definition}

Now we are ready to introduce some asymptotic behavior of solutions of (\ref{rotated}).
\begin{lemma}[\protect{\cite[\S 7.4]{Hille}}]\label{gen_pro}
Let $v$ be a non-constant solution of (\ref{rotated}) (with no boundary conditions imposed). Then the following hold.
\begin{enumerate}
\item[(i)] In each Stokes sector $S_j$, the solution $v$ is asymptotic to 
\begin{equation} \label{asymp-formula}
(const.)z^{-\frac{m}{4}}\exp\left[\pm\frac{2}{m+2}z^{\frac{m+2}{2}}\right], \quad\text{as $z \rightarrow \infty$ in every closed subsector of $S_j$.}
\end{equation}
\item[(ii)] If $v$ decays to zero in $S_j$, for some $j\in \Z$, then it must blow up in $S_{j-1}$ and $S_{j+1}$. (However, it is possible for $v$ to blow up in many adjacent Stokes sectors.) Moreover, the asymptotic expression (\ref{asymp-formula}) is valid with the same constant for the three consecutive Stokes sectors $S_{j-1}\cup cl\left(S_j\right)\cup S_{j+1},$ where $cl\left(S_j\right)$ is the closure of $S_j$.
\item[(iii)] For each Stokes sector $S_j$, there exists a solution of (\ref{rotated}) that decays in $S_j$, and there exists a solution of (\ref{rotated}) that blows up in $S_j$. And any solution of (\ref{rotated}) can be expressed as a linear combination of these two solutions.
\end{enumerate} 
\end{lemma}

\begin{remark}
{\rm In Lemma \ref{gen_pro}, we state some asymptotic behavior of solutions of (\ref{rotated}). We mention that Hille \cite[\S 7.4]{Hille} studied more general equations of the form (\ref{rossi}). In the general case, the corresponding Stokes sectors are given by (\ref{gen_stokes}). With these Stokes sectors, Lemma \ref{gen_pro} (ii) holds though the asymptotic expression (\ref{asymp-formula}) becomes more complicated due to more complicated potentials. 

Also, one can see that Lemma \ref{gen_pro} (iii) implies Proposition \ref{main2} (ii). More precisely, if eigenvalues were not simple, then equation (\ref{H_peq}) would have two linearly independent solutions satisfying the boundary conditions in (\ref{bdcond}). Hence any solution of (\ref{H_peq}) could be expressed as a linear combination of these solutions. So there would be no solution of (\ref{H_peq}) that blows up in either of the two Stokes sectors containing the rays in (\ref{bdcond}). This contradicts Lemma \ref{gen_pro} (iii). Therefore, all eigenvalues are simple.
}
\end{remark}

From now on, we denote the Stokes sectors $S_j$ as the Stokes sectors of (\ref{rotated}).
\subsection*{Admissible boundary conditions}
Notice, in particular, that 
the asymptotic expression (\ref{asymp-formula}) implies  that for each $j$, $v(z)$ either decays to zero or blows up exponentially, as $z$ approaches infinity in  closed subsectors of $S_j$. Also,  the asymptotic expression (\ref{asymp-formula}) implies that if $v(z)\rightarrow 0$ as $z\rightarrow \infty$ along one ray in $S_j$, then $v(z)\rightarrow 0$ as $z\rightarrow \infty$ along every ray in $S_j$. Likewise, if $v(z)\rightarrow \infty$ as $z\rightarrow \infty$ along one ray in $S_j$, then $v(z)\rightarrow \infty$ as $z\rightarrow \infty$ along every ray in $S_j$. 

Let $u$ be an eigenfunction of $H_{\ell}$. Then
the above observation  shows that when $\ell=2n-1$ is odd, the boundary condition (\ref{bdcond}) for $H_{\ell}$ is equivalent to having $v(z)=u(-iz)$ decaying to zero as $z\rightarrow\infty$ along rays in $S_{-n}$ and  $S_{n}$ (note that the rays $\arg z= \pm \frac{\ell+1}{m+2}\pi$ are center rays of $S_{-n}$ and  $S_{n}$). Also, when $\ell=2n$ is even, the boundary condition (\ref{bdcond}) for $H_{\ell}$ is equivalent to having $w(z)=u(-i(\omega^{\frac{1}{2}}z))$ decaying to zero as $z\rightarrow\infty$ along rays in $S_{-n-1}$ and  $S_{n}$.

As we saw above, one need not choose the two rays being symmetric, as in (\ref{bdcond}), so long as they stay in the Stokes sectors that are symmetric with respect to the imaginary axis. 
Next we explain why every solution of (\ref{rotated}) decays to zero algebraically as $z$ tends to infinity along the critical rays. To this end, we examine the asymptotic expression (\ref{asymp-formula}). Certainly, one can check $\Re \left(z^{\frac{m+2}{2}}\right)=0$ for all $z$ on the critical rays, and hence it is not difficult to see that every solution of (\ref{rotated}) decays to zero algebraically as $z$ tends to infinity along the critical rays. (Incidently, the Stokes sectors $S_j$ are the sectors where  $\Re \left( z^{\frac{m+2}{2}}\right)$ keeps a constant sign.)

One might wonder why we do not consider the eigenvalue problem
$$\left[-\frac{d^2}{dz^2}-(-1)^{\ell} (iz)^m\right]u(z)=\lambda u(z), $$
under the boundary condition (\ref{bdcond}). (Here we have the opposite sign, compared to (\ref{H_peq}), in front of $(-1)^{\ell}$.) In this case, under the rotation $v(z)=u(-iz)$, and $w(z)=v(\omega^{\frac{1}{2}}z)$ if necessary, we see that the two rays in (\ref{bdcond}) map to two of the critical rays of (\ref{rotated}). So we have algebraic decay of the solution. Thus if we require the eigenfunction  be decaying to zero `exponentially', there are no eigenvalues, while  if we require the eigenfunction  be decaying to zero merely algebraically, then every complex number $\lambda$ is an eigenvalue.  And hence we have no interest in this case.

Now we are ready to explain how Theorem \ref{main1} covers every symmetric rapid decay boundary condition. When $\ell$ is odd, the equation (\ref{H_peq}) becomes $-u^\dd(z)-(iz)^m u(z)=\lambda u(z)$. In this case, the negative imaginary axis is the center of a Stokes sector and the critical rays are $\arg z=-\frac{\pi}{2}\pm\frac{(2k-1)\pi}{m+2}$ for integers $1\leq k\leq \frac{m+3}{2}$. The two rays in (\ref{bdcond}) are not critical rays and they are, in fact, centers of the Stokes sectors. When $\ell$ is increased by $2$, the rays in (\ref{bdcond}) move, away from the negative imaginary axis, to the centers of adjacent Stokes sectors. So Theorem \ref{main1} covers all symmetric rapid decay boundary condition for the potential $-(iz)^m$. Similarly, one can see that when $\ell$ is even,  Theorem \ref{main1} covers all symmetric rapid decay boundary condition for the potential $(iz)^m$.

So far in this subsection, we have discussed all possible symmetric decaying boundary conditions, except the ``imaginary axis'' boundary condition discussed at the end of the Introduction. Next we briefly mention non-symmetric decaying boundary conditions. Let us consider, as an example, 
$$\left[-\frac{d^2}{dz^2}-(iz)^m\right]u(z)=\lambda u(z), $$
under the boundary condition that $u(z)$ decays to zero exponentially as $z$ tends to infinity along the rays $\arg z=-\frac{\pi}{2}+\frac{2\pi}{m+2}$ and $\arg z=-\frac{\pi}{2}-\frac{4\pi}{m+2}$. We set $u_1(z):=u(\omega^{-\frac{1}{2}} z)$, and then $u_1(z)$ solves
$$\left[-\frac{d^2}{dz^2}+(iz)^m\right]u_1(z)= \omega^{-1}\lambda u_1(z), $$
under the boundary condition (\ref{bdcond}) with $\ell=2$. Then
by the results in this paper, Theorem \ref{main1}, one can see that $\omega^{-1}\lambda$ is positive real. Hence $\lambda$ is not real.
In general, if we impose a decaying boundary condition along the two rays in some Stokes sectors that are not symmetric with respect to the imaginary axis, then the eigenvalues are not positive real. 

Finally, we mention that the integer $\ell$ in (\ref{H_peq}) and (\ref{bdcond}) is the same as  the number of Stokes sectors  between the two sectors where we impose the boundary condition (\ref{bdcond}).

\subsection*{Results of Sibuya} 
Next we will introduce Sibuya's results, but first we define the order of an entire function $g$ as
$$order (g)=\limsup_{r\rightarrow \infty}\frac{\log \log M(r,g)}{\log r},$$
where $M(r, g)=\max \{|g(re^{i\theta})|: 0\leq \theta\leq 2\pi\}$ for $r>0$.
If for some positive real numbers $\sigma,\, c_1,\, c_2$, we have $M(r,g)\leq c_1 \exp[c_2 r^{\sigma}]$ for all large $r$, then the order of $g$ is finite and less than or equal to $\sigma$. In this paper, we choose $\arg z^\alpha=\alpha \arg z$ for $-\pi<\arg z\leq \pi$ and $\alpha \in \R$.

Now we are ready to introduce some existence results  and asymptotic estimates of Sibuya \cite{Sibuya}. The existence of an entire solution with a specified asymptotic representation for fixed $\lambda$, is presented as well as an asymptotic expression of the value of the solution at $z=0$ as $\lambda$ tends to infinity. These results are in Theorems 6.1, 7.2, 19.1 and 20.1 of Sibuya's book \cite{Sibuya}. The following is a special case of these theorems that is enough for our argument later. 
\begin{proposition}\label{prop}
The equation (\ref{rotated}) admits a solution  $f(z,\lambda)$ with the following properties.
\begin{enumerate}
\item[(i)] $f(z,\lambda)$ is an entire function of $(z,\lambda)$.
\item[(ii)] $f(z,\lambda)$ and $f^\d(z,\lambda)=\frac{d}{dz}f(z,\lambda)$ admit the following asymptotic expressions. Let $\epsilon>0$. Then
\begin{eqnarray}
f(z,\lambda)&=&\,\, z^{-\frac{m}{4}}(1+O(z^{-1/2}))\exp\left[-\frac{2}{m+2}z^{\frac{m+2}{2}}\right],\nonumber\\
f^\d(z,\lambda)&=&-z^{\frac{m}{4}}(1+O(z^{-1/2}))\exp\left[-\frac{2}{m+2}z^{\frac{m+2}{2}} \right],\nonumber
\end{eqnarray}
as $z$ tends to infinity in  the sector $|\arg z|\leq \frac{3\pi}{m+2}-\epsilon$, uniformly on each compact set of the complex $\lambda$-plane. 
\item[(iii)] Properties \textup{(i)} and \textup{(ii)} uniquely determine the solution $f(z,\lambda)$ of (\ref{rotated}).
\item[(iv)] For each fixed $\delta>0$, $f$ and $f^\d$ also admit the asymptotic expressions, 
\begin{eqnarray}
f(0,\lambda)&=&\,\,[1+o(1)]\lambda^{-1/4}\exp\left[K\lambda^{\frac{1}{2}+\frac{1}{m}}\right],\label{eq1}\\
f^\d(0,\lambda)&=&-[1+o(1)]\lambda^{1/4}\exp\left[K\lambda^{\frac{1}{2}+\frac{1}{m}}\right],\label{eq2}
\end{eqnarray}
 as $\lambda$ tends to infinity in the sector $|\arg \lambda|\leq \pi-\delta$, where 
\begin{equation}\label{def_K}
 K=\int_0^{\infty}\left(\sqrt{1+t^m}-\sqrt{t^m}\right)\, dt.
\end{equation}
\item[(v)] The entire functions  $\lambda\mapsto f(0,\lambda)$ and $\lambda\mapsto f^\d(0,\lambda)$ have orders $\frac{1}{2}+\frac{1}{m}$.
\end{enumerate}
\end{proposition}
\begin{proof}
In  Sibuya's book \cite{Sibuya}, see Theorem 6.1 for a proof of (i) and (ii); Theorem 7.2 for a proof of (iii); and Theorem 19.1 for a proof of (iv).  And (v) is a consequence of (iv) along with Theorem 20.1. Note that properties (i), (ii) and (iv) are summarized on pages 112--113 of Sibuya's book. 
\end{proof}

The next thing we want to introduce is the Stokes multiplier. 
Let $f(z,\lambda)$ be the function in Proposition \ref{prop}. Note that $f(z,\lambda)$ decays to zero exponentially as $z\rightarrow \infty$ in $S_0$, and  blows up in $S_{-1}\cup S_1$. Then one can see that the function
\begin{equation}\label{f_kdef}
f_k(z,\lambda):=f(\omega^{-k}z,\omega^{-mk}\lambda),
\end{equation}
 which is obtained by rotating $f(z,\omega^{-mk}\lambda)$ in the $z$-variable solves (\ref{rotated}). It is clear that $f_0(z,\lambda)=f(z,\lambda)$, and that $f_k(z,\lambda)$ decays in $S_k$ and blows up in $S_{k-1}\cup S_{k+1}$ since $f(z,\omega^{-mk}\lambda)$ decays in $S_0$. Then since no non-constant solution decays in two consecutive Stokes sectors, $f_{k}$ and $f_{k+1}$ are linearly independent and hence any solution of (\ref{rotated}) can be expressed as a linear combination of these two. Especially, for some coefficients $C_{j,k}(\lambda)$ and $D_{j,k}(\lambda)$,
\begin{equation}\label{stokes}
f_{j}(z,\lambda)=C_{j,k}(\lambda)f_k(z,\lambda)+D_{j,k}(\lambda)f_{k+1}(z,\lambda),\quad j,\,k\in \Z.
\end{equation}
These $C_{j,k}(\lambda)$ and $D_{j,k}(\lambda)$ are called {\it the Stokes multipliers of $f_{j}$ with respect to $f_k$ and $f_{k+1}$}.

We then see that 
\begin{equation}\label{CD_def}
C_{j,k}(\lambda)=\frac{W_{j,k+1}(\lambda)}{W_{k,k+1}(\lambda)}\quad\text{and}\quad D_{j,k}(\lambda)=-\frac{W_{j,k}(\lambda)}{W_{k,k+1}(\lambda)},
\end{equation}
where $W_{j,k}=f_jf_k^\d -f_j^\d f_k$ is the Wronskian of $f_j$ and $f_k$. Since both $f_j$ and $f_k$ are solutions of the same linear equation (\ref{rotated}), we know that the Wronskians are constant functions of $z$. Since $f_k$ and $f_{k+1}$ are linearly independent, $W_{k,k+1}\not=0$ for all $k\in \Z$. In the next lemma, we will show that the Wronskian $W_{k,k+1}(\lambda)$ is constant, which is needed in the proof of our main theorem. 

\begin{lemma}\label{unit}
For each $k\in \Z$, the Wronskian $W_{k,k+1}(\lambda)=-2\omega^{-\frac{m}{4}-k-1}$, which is independent of $\lambda$. 
\end{lemma}
\begin{proof}
Since $f_k(z,\,\lambda)=f(\omega^{-k}z,\omega^{-mk}\lambda)$, we get 
\begin{eqnarray}
 f_{k+1}(z,\lambda)
&=&f(\omega^{-(k+1)}z,\omega^{-m(k+1)}\lambda)\nonumber\\
&=&f_k(\omega^{-1}z,\omega^{-m}\lambda)\nonumber\\
&=&f_k(\omega^{-1}z,\omega^{2}\lambda),\nonumber
\end{eqnarray}
using $\omega^{m+2}=1$. Also we see that
$$f_{k+1}^\d(z,\lambda)=\omega^{-1}f_k^\d(\omega^{-1}z,\omega^{2}\lambda).$$
Thus 
\begin{equation}\label{kplus1}
W_{j+1,k+1}(\lambda)=\omega^{-1}W_{j,k}(\omega^2\lambda).
\end{equation}

Next we compute
\begin{align}
W_{-1,0}(\lambda)&=f_{-1}f^\d_0-f_0f^\d_{-1}\nonumber\\
&=f(\omega z,\omega^{-2}\lambda)f^\d(z,\lambda)-f(z,\lambda)\omega f^\d(\omega z,\omega^{-2}\lambda).\nonumber
\end{align}
Thus as $z$ tends to infinity in $S_0$ for which above asymptotics are valid, we have
\begin{align}
W_{-1,0}(\lambda)&=-(\omega z)^{-\frac{m}{4}}z^{\frac{m}{4}}(1+O(z^{-\frac{1}{2}}))\exp\left[-\frac{2}{m+2}(\omega z)^{\frac{m+2}{2}}- \frac{2}{m+2} z^{\frac{m+2}{2}}\right]\nonumber\\
&+ z^{-\frac{m}{4}}\omega(\omega z)^{\frac{m}{4}}(1+O(z^{-\frac{1}{2}}))\exp\left[-\frac{2}{m+2}(\omega z)^{\frac{m+2}{2}}- \frac{2}{m+2} z^{\frac{m+2}{2}}\right]\nonumber\\
&=-2\omega^{-\frac{m}{4}}(1+O(z^{-\frac{1}{2}}))\qquad\text{since $\omega^{\frac{m+2}{2}}=-1.$}\nonumber
\end{align}
  Finally we see that 
\begin{equation}\label{W-10_def}
W_{-1,0}(a,\lambda)=-2\omega^{-\frac{m}{4}},
\end{equation}
since $W_{j,k}$ is independent of $z$. Thus (\ref{kplus1}) and (\ref{W-10_def}) complete the proof.

\end{proof}
In the next lemma, we will show that the Wronskians $W_{k,0}(\lambda)$ for $2\leq k\leq m$ have orders   $\frac{1}{2}+\frac{1}{m}$, which is needed in  establishing the induction step in Theorem \ref{main} and proving  the existence of eigenvalues under various boundary conditions.  This lemma is due to Sibuya \cite{Sibuya}, but we give a full proof here.
\begin{lemma}\label{unit_order}
For each $2\leq k \leq m$, the Wronskian $W_{k,0}(\lambda)$ has its order $\frac{1}{2}+\frac{1}{m}\in \left(\frac{1}{2},\,1\right)$.
\end{lemma}
\begin{proof}
First we look at 
\begin{align}
W_{k,0}(\lambda)&=f_{k}(z,\lambda)f^\d_0(z,\lambda)-f_0(z,\lambda)f^\d_{k}(z,\lambda)\nonumber\\
&=f(0,\omega^{2k}\lambda)f^\d(0,\lambda)-f(0,\lambda)\omega^{-k} f^\d(0,\omega^{2k}\lambda),\label{order_w}
\end{align}
since the Wronskian is independent of $z$ and so we can take $z=0$. We then know that  the Wronskian $W_{k,0}(\lambda)$ has order less than or equal to $\frac{1}{2}+\frac{1}{m}$ because by Lemma \ref{prop} (v) each term in the right hand side of (\ref{order_w}) has order $\frac{1}{2}+\frac{1}{m}$. So in order to show $W_{k,0}(\lambda)$ has order equal to $\frac{1}{2}+\frac{1}{m}$, it suffices to show that there exist $c_1>0$ and $c_2>0$ such that $|W_{k,0}(\lambda)|\geq c_1\exp\left[c_2|\lambda|^{\frac{1}{2}+\frac{1}{m}}\right]$ for all large $|\lambda|$ along some ray.

Next we examine the right hand side of (\ref{order_w}) along the ray 
$$\arg \lambda\equiv \theta=\pi-\frac{4\pi}{m+2},$$ 
by using the asymptotic expressions (\ref{eq1}) and (\ref{eq2}).
Notice $\theta\in \left(-\pi+\delta,\pi-\delta\right)$, for all $0<\delta<\frac{\pi}{m+2}$.
 Recall that the expressions (\ref{eq1}) and (\ref{eq2}) are defined for $\arg \lambda \in (-\pi+\delta,\pi-\delta)$. So in using (\ref{eq1}) and (\ref{eq2}) for  $f(0,\omega^{2k}\lambda)$ and $f^\d(0,\omega^{2k}\lambda)$, we will choose
\begin{equation}\label{theta_*_sign}
\arg (\omega^{2k}\lambda)\equiv \theta_*=\theta+\frac{4k\pi}{m+2}-2\pi,\quad\text{for $2\leq k\leq \frac{m+2}{2}$},
\end{equation}
so we have
\begin{equation}\label{theta_sign}
-\pi<-\frac{m\pi}{m+2}<\theta_*<\theta<\frac{m\pi}{m+2}<\pi,\quad\text{for $2\leq k\leq \frac{m+2}{2}$}.
\end{equation}
Thus we see 
\begin{eqnarray}
f(0,\omega^{2k}\lambda)&=&\,\,[1+o(1)]|\lambda|^{-\frac{1}{4}}e^{-i\frac{\theta_*}{4}}\exp\left[K|\lambda|^{\frac{1}{2}+\frac{1}{m}}e^{i\frac{m+2}{2m}\theta_*}\right],\nonumber\\
f^\d(0,\omega^{2k}\lambda)&=&-[1+o(1)]|\lambda|^{\frac{1}{4}}e^{i\frac{\theta_*}{4}}\exp\left[K|\lambda|^{\frac{1}{2}+\frac{1}{m}}e^{i\frac{m+2}{2m}\theta_*}\right].\nonumber
\end{eqnarray}
Hence from this along with (\ref{order_w}), we get
\begin{align}
W_{k,0}(\lambda)&=-[1+o(1)]|\lambda|^{1/4}e^{i\frac{\theta}{4}}|\lambda|^{-\frac{1}{4}}e^{-i\frac{\theta_*}{4}}\exp\left[K|\lambda|^{\frac{1}{2}+\frac{1}{m}}\left(e^{i\left(\frac{1}{2}+\frac{1}{m}\right)\theta}+e^{i\left(\frac{1}{2}+\frac{1}{m}\right)\theta_*}\right)\right]\nonumber\\
&+[1+o(1)]|\lambda|^{-1/4}e^{-i\frac{\theta}{4}}\omega^{-k}|\lambda|^{\frac{1}{4}}e^{i\frac{\theta_*}{4}}\exp\left[K|\lambda|^{\frac{1}{2}+\frac{1}{m}}\left(e^{i\left(\frac{1}{2}+\frac{1}{m}\right)\theta}+e^{i\left(\frac{1}{2}+\frac{1}{m}\right)\theta_*}\right)\right]\nonumber\\
&=\left[-e^{i\frac{\theta-\theta_*}{4}}+\omega^{-k}e^{i\frac{\theta_*-\theta}{4}} \right][1+o(1)]\exp\left[K|\lambda|^{\frac{1}{2}+\frac{1}{m}}\left(e^{i\left(\frac{1}{2}+\frac{1}{m}\right)\theta}+e^{i\left(\frac{1}{2}+\frac{1}{m}\right)\theta_*}\right)\right]\nonumber\\
&=-2i\omega^{-\frac{k}{2}}[1+o(1)]\exp\left[K|\lambda|^{\frac{1}{2}+\frac{1}{m}}\left(e^{i\left(\frac{1}{2}+\frac{1}{m}\right)\theta}+e^{i\left(\frac{1}{2}+\frac{1}{m}\right)\theta_*}\right)\right],\nonumber
\end{align}
where the last step is by $\theta-\theta_*=2\pi-\frac{4k\pi}{m+2}.$
So when $\arg \lambda=\pi-\frac{4\pi}{m+2}$, we have 
$$\Re \left[e^{i\left(\frac{1}{2}+\frac{1}{m}\right)\theta}+e^{i\left(\frac{1}{2}+\frac{1}{m}\right)\theta_*}\right]=\Re \left[\cos\left(\frac{m+2}{2m}\theta\right)+\cos\left(\frac{m+2}{2m}\theta_*\right)\right]>0,$$
where the last step is by (\ref{theta_sign}).

So for $2\leq k\leq\frac{m+2}{2}$, since $K>0$, there exists $c_2>0$ such that 
$$\left|W_{k,0}(\lambda)\right|\geq \exp\left[c_2|\lambda|^{\frac{1}{2}+\frac{1}{m}}\right]\quad\text{for all large $|\lambda|$ on the ray $\arg \lambda=\pi-\frac{4\pi}{m+2}$}.$$
Thus the order of $W_{k,0}(\lambda)$ for $2\leq k\leq \frac{m+2}{2}$ is $\frac{1}{2}+\frac{1}{m}$.  

Certainly, the  Wronskian $W_{k,0}(\lambda)$ for $2\leq k\leq\frac{m+2}{2}$ is blowing up in some other directions as well. If one find $\theta$ and $\theta_*$ satisfying (\ref{theta_sign}), the above argument will show that $W_{k,0}(\lambda)$ is blowing up along $\arg \lambda=\theta.$

Since $f_k(z,\,\lambda)=f(\omega^{-k}z,\omega^{-mk}\lambda)$, we get 
\begin{eqnarray}
 f_{k+1}(z,\lambda)
&=&f(\omega^{-(k+1)}z,\omega^{-m(k+1)}\lambda)\nonumber\\
&=&f_k(\omega^{-1}z,\omega^{-m}\lambda)\nonumber\\
&=&f_k(\omega^{-1}z,\omega^{2}\lambda),\nonumber
\end{eqnarray}
using $\omega^{m+2}=1$.

Next for $\frac{m+2}{2}<k\leq m$, one can choose $\theta=-\pi+\frac{4\pi}{m+2}$ and $\theta_*$ by (\ref{theta_*_sign}), and then follow an argument similar to the above to conclude that  the order of $W_{k,0}(\lambda)$ is $\frac{1}{2}+\frac{1}{m}$. Or one uses an index change to reduce to the case already considered. That is,
\begin{eqnarray}
W_{k,0}(\lambda)&=&W_{k-(m+2),0}(\lambda)\nonumber\\
&=&\omega^{m+2-k}W_{0,m+2-k}(\omega^{-2(m+2-k)}\lambda)\quad\text{by (\ref{kplus1})}\nonumber\\
&=&-\omega^{m+2-k}W_{m+2-k,0}(\omega^{-2(m+2-k)}\lambda).\nonumber
\end{eqnarray}
Then since $2\leq m+2-k<\frac{m+2}{2}$, we know the order of $W_{m+2-k,0}(\omega^{2(m+2-k)}\lambda)$ is $\frac{1}{2}+\frac{1}{m}$, and hence so is the order of $W_{k,0}(\lambda)$ for $\frac{m+2}{2}<k\leq m$.
This completes the proof. 
\end{proof}

\subsection*{Further results of Sibuya; Identifying the eigenvalues as zeros of  certain entire functions}
We can relate the zeros of $C_{-n,n-1}(\lambda)$ and $D_{-n,n}(\lambda)$ with the eigenvalues of $H_{\ell}$. First, we study the case when $\ell$ is odd, as follows.
\begin{lemma}\label{equiv}
Let $\ell=2n-1$ be odd, with $1\leq \ell\leq m-1$. Then the following are equivalent:
\begin{enumerate}
\item[(i)] A complex number $\lambda$ is an eigenvalue of $H_{\ell}$.
\item[(ii)]  $\lambda$ is a zero of the entire function $C_{-n,n-1}(\lambda)$. 
\item[(iii)]  $\lambda$ is a zero of the entire function $D_{-n,n}(\lambda)$.
\end{enumerate}
Moreover, the orders of the entire functions $C_{-n,n-1}(\lambda)$ and $D_{-n,n}(\lambda)$ are $\frac{1}{2}+\frac{1}{m}\in \left(\frac{1}{2},1\right).$
\end{lemma}

Hence, the eigenvalues are discrete because they are zeros of a non-constant entire function.
Note that the Stokes multipliers $C_{-n,n-1}(\lambda)$ and $D_{-n,n}(\lambda)$ are called {\it spectral determinants} or {\it Evans functions}, because their zeros are all eigenvalues of an eigenvalue problem. 
\begin{proof}
Suppose that $\lambda$ is an eigenvalue of $H_{2n-1}$ with the corresponding eigenfunction $u$. We let $v(z)=u(-iz)$, and then $v$ solves 
$$-v^\dd(z)+(z^m+\lambda) v(z)=0,$$ 
and decays in $S_{-n}\cup S_n$. 
Since $f_{-n}(z,\lambda)$ is another solution of the equation above that decays in $S_{-n}$, we see that $f_{-n}(z,\lambda)$ is a multiple of $v$. Similarly $f_n(z,\lambda)$ is a multiple of $v$. 
Then since 
\begin{equation}\label{rel_eq1}
\fbox{$f_{-n}(z,\lambda)=C_{-n,n-1}(\lambda)f_{n-1}(z,\lambda)+D_{-n,n-1}(\lambda)f_{n}(z,\lambda),$}
\end{equation}
and since $f_{n-1}(z,\lambda)$ blows up in $S_{n}$, we conclude that $C_{-n,n-1}(\lambda)=0$.

Conversely we suppose that  $C_{-n,n-1}(\lambda)=0$ for some $\lambda\in \C$. Then
from (\ref{rel_eq1}) we see that $f_{-n}(z,\lambda)$ is a constant multiple of $f_{n}(z,\lambda)$. Thus both are decaying in $S_{-n}\cup S_n$, and hence $u(z):=f_{-n}(iz,\lambda)$ is an eigenfunction of $H_{2n-1}$ with the corresponding eigenvalue $\lambda$.

Similarly, since 
\begin{equation}\nonumber
\fbox{$f_{-n}(z,\lambda)=C_{-n,n}(\lambda)f_{n}(z,\lambda)+D_{-n,n}(\lambda)f_{n+1}(z,\lambda),$}
\end{equation}
we see that $\lambda$ is an eigenvalue of $H_{2n-1}$ if and only if 
 $D_{-n,n}(\lambda)=0$. 

Finally, since $W_{k,k+1}(\lambda)$ is a constant by Lemma \ref{unit} and since $W_{k,0}(\lambda)$ for $2\leq k\leq m$ has order $\frac{1}{2}+\frac{1}{m}$ by Lemma \ref{unit_order}, we see from (\ref{CD_def}) that $C_{-n,n-1}(\lambda)$ and $D_{-n,n}(\lambda)$ are of order $\frac{1}{2}+\frac{1}{m}\in \left(\frac{1}{2},1\right).$ This completes the proof.
\end{proof}

Second, we can relate the zeros of $C_{-n-1,n-1}(\lambda)$ and $D_{-n-1,n}(\lambda)$ with the eigenvalues of $H_{\ell}$, when $\ell$ is even, as follows.
\begin{lemma}\label{equiv1}
Let $\ell=2n$ be even, with $1\leq \ell\leq m-1$. Then the following are equivalent:
\begin{enumerate}
\item[(i)] A complex number $\lambda$ is an eigenvalue of $H_{\ell}$.
\item[(ii)]  $\lambda$ is a zero of the entire function $\lambda\mapsto C_{-n,n}(\omega^{-1} \lambda)$. 
\item[(iii)]  $\lambda$ is a zero of the entire function $\lambda\mapsto D_{-n-1,n}(\omega \lambda)$.
\end{enumerate}
Moreover, the orders of the entire functions $\lambda\mapsto C_{-n,n}(\omega^{-1} \lambda)$ and $\lambda\mapsto D_{-n-1,n}(\omega \lambda)$ are $\frac{1}{2}+\frac{1}{m}\in \left(\frac{1}{2},1\right).$
\end{lemma} 
Note again that the Stokes multipliers $C_{-n,n}(\omega^{-1}\lambda)$ and $D_{-n-1,n}(\omega\lambda)$ are  spectral determinants or  Evans functions.
\begin{proof}
Let $u$ be an eigenfunction for $H_{2n}$ with the corresponding eigenvalue $\lambda$.
Then $v(z):=u(-i z)$ solves
$$v^\dd(z)+z^mv(z)=\lambda v(z).$$

Next we let $w_1(z):=v(\omega^{-\frac{1}{2}}z)$. Then $w_1(z)$ solves
\begin{equation}
-w_1^\dd(z)+\left[z^m+\omega^{-1}\lambda\right] w_1(z)=0.
\end{equation}
One can then check that the boundary condition for $H_{2n}$ becomes that
$w_1(z)\rightarrow 0$ as $z\rightarrow\infty$ in $S_{-n}\cup S_{n+1}$. Then like we did for Lemma \ref{equiv}, using 
\begin{equation}\nonumber
f_{-n}(z,\omega^{-1}\lambda)=C_{-n,n}(\omega^{-1}\lambda)f_{n}(z,\omega^{-1}\lambda)+D_{-n,n}(\omega^{-1}\lambda)f_{n+1}(z,\omega^{-1}\lambda),
\end{equation}
one can show that $\lambda$ is an eigenvalue of $H_{2n}$ if and only if $C_{-n,n}(\omega^{-1}\lambda)=0$.

Similarly,
 we let $w_2(z):=v(\omega^{\frac{1}{2}}z)$. Then $w_2(z)$ solves
\begin{equation}
-w_2^\dd(z)+\left[z^m+\omega\lambda\right] w_2(z)=0.
\end{equation}
One can then check that the boundary condition for $H_{2n}$ becomes that
$w_2(z)\rightarrow 0$ as $z\rightarrow\infty$ in $S_{-n-1}\cup S_{n}$. Then using 
\begin{equation}\nonumber
f_{-n-1}(z,\omega\lambda)=C_{-n-1,n}(\omega\lambda)f_{n}(z,\omega\lambda)+D_{-n-1,n}(\omega\lambda)f_{n+1}(z,\omega\lambda),
\end{equation}
one can show that $\lambda$ is an eigenvalue of $H_{2n}$ if and only if $D_{-n-1,n}(\omega\lambda)=0$. And this complete the proof.
\end{proof}

Next we want an infinite product representation of the entire function $\lambda\mapsto f(0,\lambda)$. Recall that the integer $m\geq 3$ is fixed.
\begin{lemma}\label{asymp}
The functions $f(0,\lambda)$ and  $f^\d(0,\lambda)$ have infinitely many zeros $E_j<0$ and $E^\d_j<0$, respectively.  And they admit the following infinite product representations:
\begin{eqnarray}
f(0,\lambda)= A_1 \prod_{j=1}^{\infty}\left(1-\frac{\lambda}{E_j}\right)\quad\text{for some  $A_1 \in \C-\{0\}$.}\label{inf_1}\\
f^\d(0,\lambda)= A_2 \prod_{j=1}^{\infty}\left(1-\frac{\lambda}{E^\d_j}\right)\quad\text{for some  $A_2 \in \C-\{0\}$.}\label{inf_2}
\end{eqnarray}
\end{lemma}
\begin{proof}
We know that $f(0,\lambda)$ and $f^\d(0,\lambda)$ have orders $\frac{1}{2}+\frac{1}{m}\in (0,1)$ by Proposition \ref{prop} (v), and hence by the Hadamard factorization theorem (see, for example, Theorem 14.2.6 on page 199 of \cite{Hille2}), we know  $f(0,\lambda)$ and $f^\d(0,\lambda)$  have infinite product representations (\ref{inf_1}) and (\ref{inf_2}) where $f(0,E_j)=0$ and $f^\d(0,E^\d_j)=0$ for all $j\in \N$. So in order to complete the proof, we need to show $E<0$ if  $f(0,E)=0$, and $E^\d<0$ if $f^\d(0,E^\d)=0$.

Suppose $f(0,E_*)=0$ or $f^\d(0,E_*)=0$.  By the definition, we know $f(z,E_*)$ solves
$$-f^\dd(z,E_*)+z^mf(z,E_*)=-E_*f(z,E_*),$$
and decays to zero exponentially in $S_0$. (Note that $-\frac{d^2}{dx^2}+x^m$ is Hermitian on the positive real axis.)
In order to show $E_*<0$, we multiply this equation by $\overline{f(z,E_*)}$ and integrate over the positive real axis to get
$$-\int_0^{\infty} f^\dd(x,E_*)\overline{f(x,E_*)}\, dx+\int_0^{\infty} x^m|f(x,E_*)|^2\, dx=-E_*\int_0^{\infty} |f(x,E_*)|^2\, dx.$$
Next we integrate the first term by parts, and use $f(0,E_*)=0$ or $f^\d(0,E_*)=0$. Then clearly the left hand-side of the resulting equation is positive, and hence we conclude $E_*<0$. This completes the proof.
\end{proof}
Next, we will prove a symmetry lemma, regarding $f(z,\lambda)$. 
\begin{lemma}
For each $\lambda\in \C$, 
\begin{equation}\label{eq111}
\overline{f(\overline{z},\lambda)}=f(z,\overline{\lambda})\quad\text{and}\quad \overline{f^\d(\overline{z},\lambda)}=f^\d(z,\overline{\lambda}).
\end{equation}
\end{lemma}
\begin{proof}
This lemma for $z=0$ is contained in Lemma 8 in \cite{Shin2}, and in fact that is all we will need in this paper. A proof of this lemma is essentially the same as that of \cite[Lemma 8]{Shin2}. So we omit the proof here.
\end{proof}

\section{The induction step}\label{induction_step}
In proving Theorem \ref{main1}, we will use induction on $\ell$. The induction step will be provided by the following theorem.
\begin{theorem}\label{main}
 If $1\leq \ell \leq\frac{m}{2}-1$ and all the eigenvalues of $H_{\ell}$ lie in the sector $|\arg \cdot|\leq \frac{2\pi}{m+2}$, then  every eigenvalue of  $H_{\ell+1}$ is positive real. 
\end{theorem}

\begin{proof}[Proof of Theorem \ref{main} (Case I: $\ell=2n-1$ is odd, with $1\leq n\leq \frac{m}{4}$)]
Suppose that all the eigenvalues $\sigma_j$ of $H_{2n-1}$ lie in the sector $|\arg \sigma_j|\leq\frac{2\pi}{m+2}$. That is, zeros $\sigma_j$ of the entire function $\sigma\mapsto D_{-n,n}(\sigma)$ lie in the sector $|\arg \sigma|\leq\frac{2\pi}{m+2}$ (see Lemma \ref{equiv}). Then we want to show that each eigenvalue $\lambda$ of $H_{2n}$ is positive real. 

Suppose $u(z)$ is an eigenfunction of $H_{2n}$ with the eigenvalue $\lambda$. That is, $u(z)$ solves 
$$-u^\dd(z)+(iz)^mu(z)=\lambda u(z),$$
and decays to zero as $z$ tends to infinity along the two rays $\arg z=-\frac{\pi}{2}\pm (2n+1)\frac{\pi}{m+2}.$ We then let $v(z)=u(-iz)$, and so $v(z)$ solves
\begin{equation}\label{v_eq}
v^\dd(z)+z^m v(z)=\lambda v(z),
\end{equation}
and decays  to zero as $z$ tends to infinity along the two rays $\arg z=\pm (2n+1)\frac{\pi}{m+2}.$

Next we let $w(z)=v(\omega^{-\frac{1}{2}} z)$ so that $w(z)$ solves  
\begin{equation}\label{w_eq}
-w^\dd(z)+\left[z^m+\omega^{-1} \lambda\right]w(z)=0,
\end{equation}
and decays to zero as $z$ tends to infinity along the two rays $\arg z=(2n+2)\frac{\pi}{m+2},\,-2n\frac{\pi}{m+2}$ that are the center rays of $S_{n+1}$ and  $S_{-n}$.

Then we examine
\begin{equation}\nonumber
f_{-n}(z,\omega^{-1}\lambda)=C_{-n,n}(\omega^{-1}\lambda)f_{n}(z,\omega^{-1}\lambda)+D_{-n,n}(\omega^{-1}\lambda)f_{n+1}(z,\omega^{-1}\lambda),
\end{equation}
or equivalently,
\begin{equation}\label{pdexp}
f(\omega^n z,\omega^{-2n-1}\lambda)=C_{-n,n}(\omega^{-1}\lambda)f(\omega^{-n}z,\omega^{2n-1}\lambda)+D_{-n,n}(\omega^{-1}\lambda)f(\omega^{-n-1}z,\omega^{2n+1}\lambda).
\end{equation}
We see that $C_{-n,n}(\omega^{-1}\lambda)=0$ by Lemma \ref{equiv1}, and hence $D_{-n,n}(\omega^{-1}\lambda)\not=0$. 

Next we will show that $\left|D_{-n,n}(\omega^{-1}\lambda)\right|\leq 1$ when $\lambda$ is an eigenvalue of $H_{2n}$ with $\Im \lambda\geq 0$.
To this end, we evaluate the equation (\ref{pdexp}) and its differentiated form at $z=0$ to get
\begin{eqnarray}
f(0,\omega^{-2n-1}\lambda)&=&D_{-n,n}(\omega^{-1}\lambda)f(0,\omega^{2n+1}\lambda)\qquad\text{and}\label{abs1}\\
\omega^nf^\d(0,\omega^{-2n-1}\lambda)&=&\omega^{-n-1}D_{-n,n}(\omega^{-1}\lambda)f^\d(0,\omega^{2n+1}\lambda).\label{abs10}
\end{eqnarray}
Then since $\lambda$ is an eigenvalue, so is $\overline{\lambda}$. Thus we have 
\begin{eqnarray}
f(0,\omega^{-2n-1}\overline{\lambda})&=&D_{-n,n}(\omega^{-1}\overline{\lambda})f(0,\omega^{2n+1}\overline{\lambda})\qquad\text{and}\nonumber\\
\omega^nf^\d(0,\omega^{-2n-1}\overline{\lambda})&=&\omega^{-n-1}D_{-n,n}(\omega^{-1}\overline{\lambda})f^\d(0,\omega^{2n+1}\overline{\lambda}).\nonumber
\end{eqnarray}
Then we take the complex conjugates of these, and apply (\ref{eq111}) at $z=0$ to get
\begin{eqnarray}
f(0,\omega^{2n+1}\lambda)&=&\overline{D_{-n,n}(\omega^{-1}\overline{\lambda})}f(0,\omega^{-2n-1}\lambda)\qquad\text{and}\nonumber\\
\omega^{-n}f^\d(0,\omega^{2n+1}\lambda)&=&\omega^{n+1}\overline{D_{-n,n}(\omega^{-1}\overline{\lambda})}f^\d(0,\omega^{-2n-1}\lambda).\nonumber
\end{eqnarray}
So these along with (\ref{abs1}) and (\ref{abs10}) imply
\begin{equation}\label{D_unit1}
D_{-n,n}(\omega^{-1}\lambda)\overline{D_{-n,n}(\omega^{-1}\overline{\lambda})}=1.
\end{equation}

Clearly the order of the entire function $\sigma\mapsto D_{-n,n}(\sigma)$ is $\frac{1}{2}+\frac{1}{m}\in (0,1)$. So by the Hadamard factorization theorem we have 
$$D_{-n,n}(\lambda)=B\prod_{j=1}^{\infty}\left(1-\frac{\lambda}{\sigma_j}\right),\quad\text{for some $B\in \C-\{0\}$},$$
where the $\sigma_j$ are the zeros of $D_{-n,n}(\sigma)$, and so $|\arg \sigma_j|\leq \frac{2\pi}{m+2}$ for all $j\in \N$, by hypothesis.

Thus 
\begin{equation}\label{prod_eq}
\left|\frac{D_{-n,n}(\omega^{-1}\lambda)}{D_{-n,n}(\omega^{-1}\overline{\lambda})}\right|=\prod_{j=1}^{\infty}\left|\frac{1-\frac{\omega^{-1}\lambda}{\sigma_j}}{1-\frac{\omega^{-1}\overline{\lambda}}{\sigma_j }}\right|=\prod_{j=1}^{\infty}\left|\frac{\omega \sigma_j -\lambda}{\omega\sigma_j -\overline{\lambda}}\right|\leq 1,\quad\text{when $\Im \lambda\geq 0$,}
\end{equation}
since $\Im(\omega \sigma_j)\geq 0$ for all $j\in \N$. Hence this along with (\ref{D_unit1}) implies
 $$\left|D_{-n,n}(\omega^{-1}\lambda)\right|\leq 1,\qquad\text{when $\Im \lambda\geq 0$.} $$

Since the non-constant entire function $f(z, \omega^{2n+1}\lambda)$ solves 
$$-f^\dd(z, \omega^{2n+1}\lambda)+[z^m+\omega^{2n+1}\lambda]f(z, \omega^{2n+1}\lambda)=0,$$
 we know  $f(0, \omega^{2n+1}\lambda)$ and $f^\d(0, \omega^{2n+1}\lambda)$ cannot be zero at the same time. Otherwise, $f(z, \omega^{2n+1}\lambda)= 0$ for all $z\in\C$.

Suppose that $f(0, \omega^{2n+1}\lambda)\not=0$. 
Then  we get
\begin{equation}\nonumber
1\geq \left|D_{-n,n}(\omega^{-1}\lambda)\right|=\prod_{j=1}^{\infty}\left|\frac{1-\frac{\omega^{-2n-1}\lambda}{E_j}}{1-\frac{\omega^{2n+1}\lambda}{E_j}}\right|=\prod_{j=1}^{\infty}\left|\frac{\omega^{2n+1}E_j -\lambda}{\omega^{2n+1}E_j -\overline{\lambda}}\right|\geq 1,\quad\text{when $\Im \lambda\geq 0$,}
\end{equation}
where the last inequality holds since $\Im (\omega^{2n+1}E_j)\leq 0$ if $0\leq\arg (\omega^{2n+1})\leq\pi$, by Lemma \ref{asymp} (this is where we use $1\leq n\leq\frac{m}{4}$).
So we have 
\begin{equation}\label{D_unit}
\left|D_{-n,n}(\omega^{-1}\lambda)\right|= 1,\quad\text{when $\Im \lambda\geq 0$}.
\end{equation}

Similarly, when  $f^\d(0, \omega^{2n+1}\lambda)\not=0$, we get
\begin{equation}\nonumber
1\geq \left|D_{-n,n}(\omega^{-1}\lambda)\right|=\prod_{j=1}^{\infty}\left|\frac{1-\frac{\omega^{-2n-1}\lambda}{E_j^\d}}{1-\frac{\omega^{2n+1}\lambda}{E_j^\d}}\right|=\prod_{j=1}^{\infty}\left|\frac{\omega^{2n+1}E_j^\d -\lambda}{\omega^{2n+1}E_j^\d -\overline{\lambda}}\right|\geq 1,\quad\text{when $\Im \lambda\geq 0$,}
\end{equation}
where the $E^\d_j$ are zeros of $f^\d(0,E)$. So we again have (\ref{D_unit}). Hence (\ref{prod_eq}) along with combining (\ref{D_unit1}) and (\ref{D_unit}) gives 
\begin{equation}\label{D_unit2}
\prod_{j=1}^{\infty}\left|\frac{\omega \sigma_j -\lambda}{\omega\sigma_j -\overline{\lambda}}\right|= 1.
\end{equation}
Since $|\arg \sigma_j|\leq \frac{2\pi}{m+2}$ for all $j\in\N$, we know $\Im \left(\omega \sigma_j\right)\geq 0$ for all $j\in \N$.
Moreover, $\Im \left(\omega \sigma_j\right)> 0$ for some $j$ since both $\sigma_j$ and $\overline{\sigma_j}$ are eigenvalues of the $\mathcal{PT}$-symmetric oscillator $H_{2n-1}$.
Therefore, from (\ref{D_unit2}) we conclude $\lambda=\overline{\lambda}$, and so $\lambda$ is real.

We still need to show positivity of the eigenvalues.
The function $v(z)$ solves (\ref{v_eq}) and we know $\lambda\in\R$. Also, one can check that $\overline{v(\overline{z})}$ solves the same equation. 
Then since the eigenvalues are simple, $v(z)$ and $\overline{v(\overline{z})}$ must be linearly dependent, and hence  $v(z)=c\overline{v(\overline{z})}$ for some $c\in \C$. Since  $|v(z)|$ and $|v(\overline{z})|$ agree on the real line, we see that $|c|=1$ and so $|v(z)|=|v(\overline{z})|$ for all  $z\in \C.$ That is, $|v(x+iy)|$ is even in $y$.
From this we have that 
\begin{equation}\label{real=0}
0=\left.\frac{\partial}{\partial\,y}|v(x+iy)|^2\right|_{y=0}=-2\Im \left(v^\d(x)\overline{v(x)}\right),\quad\text{for all}\quad x\in \R.
\end{equation}

Let $g(r)=v(r e^{i\theta})$ for $2n\frac{\pi}{m+2}<\theta<(2n+2)\frac{\pi}{m+2}$.
Then $g(r)$ satisfies
$$e^{-2i\theta}g^\dd(r)+e^{m i\theta}r^mg(r)=\lambda g(r).$$
We then multiply this by $\overline{g(r)}$ and integrate over $r\in [0,\infty)$ to get
$$e^{-2i\theta}\int_0^{\infty}g^\dd(r)\overline{g(r)}\, dr+e^{m i\theta}\int_0^{\infty}r^m|g(r)|^2\, dr=\lambda\int_0^{\infty} |g(r)|^2\, dr.$$
Next we integrate the first term by parts  and multiply the resulting equation by $e^{i\theta}$ to get 
$$-e^{-i\theta}g^\d(0)\overline{g(0)}-e^{-i\theta}\int_0^{\infty}|g^\d(r)|^2\, dr+e^{(m+1) i\theta}\int_0^{\infty}r^m|g(r)|^2\, dr=\lambda e^{i\theta}\int_0^{\infty} |g(r)|^2\, dr.$$
The we use $e^{-i\theta}g^\d(0)=v(0)$ and take the imaginary parts to get, for all $\frac{2n\pi}{m+2}<\theta<\frac{(2n+2)\pi}{m+2}$,
$$\sin\theta \int_0^{\infty}|g^\d(r)|^2\, dr+\sin[(m+1)\theta]\int_0^{\infty}r^m|g(r)|^2\, dr=\lambda \sin\theta\int_0^{\infty} |g(r)|^2\, dr.$$
We evaluate this at $\theta=\frac{(2n+1)\pi}{m+1}$ to have $\lambda>0$. Note that since $1\leq n\leq\frac{m-2}{4}$, we see that $\frac{2n\pi}{m+2}<\theta=\frac{(2n+1)\pi}{m+1}<\frac{(2n+2)\pi}{m+2}$.
\end{proof}

\begin{proof}[Proof of Theorem \ref{main} (Case II: $\ell=2n$ is even,  with $1\leq n\leq\frac{m-2}{4}$)]
Suppose that all the eigenvalues $\tau_j$ of $H_{2n}$ lie in the sector $|\arg \tau|\leq\frac{2\pi}{m+2}$. That is, zeros of the entire function $\tau\mapsto D_{-n-1,n}(\omega\tau)$ lie in the sector $|\arg \tau|\leq\frac{2\pi}{m+2}$. Then we want to show that each eigenvalue $\lambda$ of $H_{2n+1}$ is positive real. 

First, we examine $D_{-n-1,n}(\tau)$. From Lemma \ref{equiv1} (iii), we know that the zeros of $\tau\mapsto D_{-n-1,n}(\omega\tau)$ are the eigenvalues $\tau_j$ of $H_{2n}$, which lie in the sector $|\arg \tau|\leq \frac{2\pi}{m+2}$, by hypothesis. So we have
$$D_{-n-1,n}(\omega\tau)=B_1\prod_{j=1}^{\infty}\left(1-\frac{\tau}{\tau_j}\right),\quad\text{for some $B_1\in \C-\{0\}$,}$$
where $\Im (\omega\tau_j)\geq 0$ for all $j\in \N$. Hence
$$D_{-n-1,n}(\lambda)=B_1\prod_{j=1}^{\infty}\left(1-\frac{\omega^{-1}\lambda}{\tau_j}\right).$$
Thus
\begin{equation}\label{D_unit4}
\left|\frac{D_{-n-1,n}(\lambda)}{D_{-n-1,n}(\overline{\lambda})}\right|=\prod_{j=1}^{\infty}\left|\frac{1-\frac{\omega^{-1}\lambda}{\tau_j}}{1-\frac{\omega^{-1}\overline{\lambda}}{\tau_j}}\right|=\prod_{j=1}^{\infty}\left|\frac{\omega \tau_j -\lambda}{\omega\tau_j -\overline{\lambda}}\right|\leq 1,\quad\text{when $\Im \lambda\geq 0$,}
\end{equation}
since $\Im(\omega \tau_j)\geq 0$. 

Next, we suppose that $u(z)$ is an eigenfunction of $H_{2n+1}$ with the eigenvalue $\lambda$. That is, $u(z)$ solves 
$$-u^\dd(z)+(-1)^{2n+1}(iz)^mu(z)=\lambda u(z),$$
and decays to zero as $z$ tends to infinity along the two rays $\arg z=-\frac{\pi}{2}\pm (2n+2)\frac{\pi}{m+2}.$ We then let $v(z)=u(-iz)$, and so $v(z)$ solves
\begin{equation}\label{v_eq1}
-v^\dd(z)+(z^m+\lambda) v(z)=0,
\end{equation}
and decays  to zero as $z$ tends to infinity along the two rays $\arg z=\pm (2n+2)\frac{\pi}{m+2}$ that are the center rays of $S_{n+1}$ and  $S_{-n-1}$. 

Then we examine
\begin{equation}\label{pdexp1}
f_{-n-1}(z,\lambda)=C_{-n-1,n}(\lambda)f_{n}(z,\lambda)+D_{-n-1,n}(\lambda)f_{n+1}(z,\lambda).
\end{equation}
We see that $C_{-n-1,n}(\lambda)=0$ by Lemma \ref{equiv}, and hence $D_{-n-1,n}(\lambda)\not=0$. 
So we have 
\begin{equation}\label{lambda_1}
f(\omega^{n+1}z,\omega^{-2n-2}\lambda)=D_{-n-1,n}(\lambda)f(\omega^{-n-1}z,\omega^{2n+2}\lambda).
\end{equation}

Then we will show that $\left|D_{-n-1,n}(\lambda)\right|\leq 1$ when $\lambda$ is an eigenvalue of $H_{2n+1}$ with $\Im \lambda\geq 0$. Suppose $f(0,\omega^{-2n-2}\lambda)\not=0$. 
Since $\lambda$ is an eigenvalue, so is $\overline{\lambda}$. Thus we replace $\lambda$ by $\overline{\lambda}$ in (\ref{lambda_1}), and then evaluate the resulting equation at $z=0$ to get
$$ f(0,\omega^{-2n-2}\overline{\lambda})=D_{-n-1,n}(\overline{\lambda})f(0,\omega^{2n+2}\overline{\lambda}).$$ 
Then we take the complex conjugate of this and apply (\ref{eq111}) so that we have 
$$f(0,\omega^{2n+2}\lambda)=\overline{D_{-n-1,n}(\overline{\lambda})}f(0,\omega^{-2n-2}\lambda).$$Combining this with (\ref{lambda_1}) at $z=0$ gives
\begin{equation}\label{D=1}
\overline{D_{-n-1,n}(\overline{\lambda})}D_{-n-1,n}(\lambda)=1,
\end{equation}
since $f(0,\omega^{-2n-2}\lambda)\not=0$.

Similarly, when $f^\d(0,\omega^{-2n-2}\lambda)\not=0$, we get (\ref{D=1}) again.

The equation (\ref{D=1}) along with the inequality in (\ref{D_unit4}) implies
$\left|D_{-n-1,n}(\lambda)\right|\leq 1$ when $\lambda$ is an eigenvalue of $H_{2n+1}$ with $\Im \lambda\geq 0$.
So we get from (\ref{lambda_1}) at $z=0$
\begin{equation}\label{D_unit21}
1\geq \left|D_{-n-1,n}(\lambda)\right|=\prod_{j=1}^{\infty}\left|\frac{1-\frac{\omega^{-2n-2}\lambda}{E_j}}{1-\frac{\omega^{2n+2}\lambda}{E_j}}\right|=\prod_{j=1}^{\infty}\left|\frac{\omega^{2n+2}E_j -\lambda}{\omega^{2n+2}E_j -\overline{\lambda}}\right|\geq 1,\quad\text{when $\Im \lambda\geq 0$,}
\end{equation}
where the last inequality holds since $\Im (\omega^{2n+2}E_j)\leq 0$ if $0\leq\arg (\omega^{2n+2})\leq\pi$ (this is where we use $1\leq n\leq\frac{m-2}{4}$).
Then like in the proof for the case $\ell$ odd, we have 
$\left|D_{-n-1,n}(\lambda)\right|=1,$
which is also obtained when $f(0,\omega^{-2n-2}\lambda)=0$ (and hence $f^\d(0,\omega^{-2n-2}\lambda)\not=0$.)
Therefore, we conclude $\lambda=\overline{\lambda}$, and so $\lambda$ is real,  like in the proof of the case $\ell$ odd.

We still need to show positivity of the eigenvalues $\lambda$. Recall that $\lambda\in\R$, and
the function $v(z)=u(-iz)$ solves (\ref{v_eq1}).  
Let $g(r)=v(r e^{i\theta})$. 
Then $g(r)$ satisfies
$$e^{-2i\theta}g^\dd(r)+e^{m i\theta}r^mg(r)=\lambda g(r).$$
We  multiply this by $\overline{g(r)}$ and integrate over $r\in [0,\infty)$ to get
$$e^{-2i\theta}\int_0^{\infty}g^\dd(r)\overline{g(r)}\, dr+e^{m i\theta}\int_0^{\infty}r^m|g(r)|^2\, dr=\lambda\int_0^{\infty} |g(r)|^2\, dr.$$
Since $v(z)$  decays exponentially to zero as $z$ tends to infinity in $S_{n+1}\cup S_{-n-1}$, we have integratibility  for $(2n+1)\frac{\pi}{m+2}<\theta<(2n+3)\frac{\pi}{m+2}$.

Note that since $\lambda\in\R$, the equation (\ref{real=0}) is valid for this case as well.
We integrate the first term by parts  and multiply the resulting equation by $e^{i\theta}$ to get 
$$-e^{-i\theta}g^\d(0)\overline{g(0)}-e^{-i\theta}\int_0^{\infty}|g^\d(r)|^2\, dr+e^{(m+1) i\theta}\int_0^{\infty}r^m|g(r)|^2\, dr=\lambda e^{i\theta}\int_0^{\infty} |g(r)|^2\, dr.$$
Then we use $e^{-i\theta}g^\d(0)=v(0)$ and take the imaginary parts with using (\ref{real=0}) to get, for all $\frac{(2n+1)\pi}{m+2}<\theta<\frac{(2n+3)\pi}{m+2}$,
$$\sin\theta \int_0^{\infty}|g^\d(r)|^2\, dr+\sin[(m+1)\theta]\int_0^{\infty}r^m|g(r)|^2\, dr=\lambda \sin\theta\int_0^{\infty} |g(r)|^2\, dr.$$

We evaluate this at $\theta=\frac{(2n+1)\pi}{m+1}$ to have $\lambda>0$. Note that since $1\leq n\leq\frac{m-2}{4}$, we see that $\frac{(2n+1)\pi}{m+2}<\theta=\frac{(2n+1)\pi}{m+1}<\frac{(2n+3)\pi}{m+2}$.
\end{proof}

\section{Proof of Theorem \ref{main1}}\label{main_proof}
In proving Theorem \ref{main1}, our induction basis is provided by the following lemma that is due to Dorey et al. \cite{Dorey2} (see also \cite[Theorem 2]{Shin2}).  
\begin{lemma}\label{H_1_lemma}
The eigenvalues $\lambda$ of $H_1$ are all positive real.
\end{lemma}
Here we will give an outline of the proof.
\begin{proof}
Let $\lambda$ be an eigenvalue of $H_1$ with the corresponding eigenfunction $u(z)$. Then we set $v(z)=u(-iz)$, and hence $v(z)$ solves
$$-v(z)+\left(z^m+\lambda\right) v(z),$$
and $v(z)$ decays in $S_{-1}\cup S_1$.

Then we consider
\begin{equation}\label{base_eq}
f_{-1}(z,\lambda)=C_{-1,0}(\lambda)f_0(z,\lambda)+D_{-1,0}(\lambda)f_1(z,\lambda).
\end{equation}
So we see that $C_{-1,0}(\lambda)=0$. Moreover, we know that from (\ref{CD_def}) and (\ref{kplus1}), and Lemma \ref{unit}, 
$$\left|D_{-1,0}(\lambda)\right|=\left|\frac{W_{-1,0}(\lambda)}{W_{0,1}(\lambda)}\right|=1.$$

Next, we use infinite product forms of either (\ref{base_eq}) when $f_{-1}(0,\lambda)\not=0$, or its differentiated form when $f_{-1}^\d(0,\lambda)\not=0$ at $z=0$. Then like in the proof of Theorem \ref{main}, one can show that $\lambda>0$, since the hypothesis in Theorem \ref{main} is needed for showing  $\left|D_{-1,0}(\lambda)\right|=1$ only.
\end{proof}
Now we are ready to prove Theorem \ref{main1}.
\begin{proof}[Proof of Theorem \ref{main1}]
For integers $1\leq\ell\leq\frac{m}{2}$, the theorem easily follows from induction on $\ell$ along with Theorem \ref{main} and Lemma \ref{H_1_lemma}.

So we assume $\frac{m}{2}<\ell\leq m-1$. Suppose $\lambda$ is an eigenvalue of $H_{\ell}$ with the corresponding eigenfunction $u_1(z)$.
Then we let $u(z):= u_1(-z)$, and hence $u(z)$ solves
$-u^\dd(z)+(-1)^{\ell}(-iz)^mu(z)=\lambda u(z).$ That is, $u(z)$ solves
\begin{equation}\nonumber
-u^\dd(z)+(-1)^{m-\ell}(iz)^mu(z)=\lambda u(z),
\end{equation}
since $(-1)^{2\ell}=1$.

Next, we examine the boundary condition.
It is clear that since $u_1(z)$ decays along the two rays $\arg z=-\frac{\pi}{2}\pm \frac{\ell+1}{m+2}\pi$, $u(z)$ decays along the two rays 
\begin{eqnarray}
\arg z&=&-\pi-\frac{\pi}{2}+ \frac{\ell+1}{m+2}\pi =-\frac{\pi}{2}-\frac{(m-\ell)+1}{m+2}\pi\quad\text{and}\nonumber\\
\arg z&=&\,\,\,\,\,\pi-\frac{\pi}{2}- \frac{\ell+1}{m+2}\pi=-\frac{\pi}{2}+\frac{(m-\ell)+1}{m+2}\pi,\nonumber
\end{eqnarray} 
which is the boundary condition (\ref{bdcond}) with $m-\ell$ in the place of $\ell$.
Hence $u(z)$ is an eigenfunction of $H_{m-\ell}$ with the corresponding eigenvalue $\lambda$.
Since $\frac{m}{2}<\ell\leq m-1$, we see that $1\leq m-\ell< \frac{m}{2}$. So by induction with help of Theorem \ref{main} and Lemma \ref{H_1_lemma}, we conclude $\lambda>0$. This completes the proof.
\end{proof}
\begin{remark}
{\rm Our method in this paper closely follows the $\ell=1$ method of Dorey et al.\ \cite{Dorey2} and the author \cite{Shin2}. One big difference is that $\ell=1$ implies $\left|D_{-1,0}(\sigma)\right|=1$ for all $\sigma\in\C$, while for $1<\ell< m-1$, the corresponding functions $\sigma\mapsto D_{-n,n}(\omega^{-1}\sigma)$ in (\ref{pdexp}) and $\sigma\mapsto D_{-n-1,n}(\sigma)$ in (\ref{pdexp1}) are entire functions of order $\frac{1}{2}+\frac{1}{m}$. However, when $\lambda$ is an eigenvalue of $H_{\ell+1}$, under some hypothesis on the eigenvalues $\sigma$ of $H_{\ell}$ we are able to show that $\left|D_{-n,n}(\omega^{-1}\lambda)\right|=1$
for $\ell=2n-1$  odd, and $\left|D_{-n-1,n}(\lambda)\right|=1$ for $\ell=2n$  even. This is the new and main idea in proving the induction step, Theorem \ref{main}.
}
\end{remark}
\section{Hopes and challanges in extending to more general potentials}\label{discussion_de}
In this section, we  mention some hopes and challenges in extending our induction methods to more general polynomial potentials.
  As  mentioned at the end of the previous section, we used the induction on $\ell$ to show that $\left|D_{-n,n}(\omega^{-1}\lambda)\right|=1$
for $\ell=2n-1$  odd, and $\left|D_{-n-1,n}(\lambda)\right|=1$ for $\ell=2n$ even.
Of course, if one can find another way of proving these identities it will help us find some new reality results of eigenvalues for  more general polynomial potentials.

In \cite{Shin2}, the author studied 
\begin{equation}\label{shin_eq}
 H_{\ell}^P u(z)=\left[-\frac{d^2}{dz^2}+(-1)^{\ell}(iz)^m+P(iz)\right]u(z)=\lambda u(z),
\end{equation}
 under the boundary condition (\ref{bdcond}) with $\ell=1$, 
where $P(z)=a_1z^{m-1}+a_2z^{m-2}+\cdots+a_{m-1}z$ is a real polynomial.
I proved that if for some $1\leq j\leq\frac{m}{2}$ we have $(j-k)a_k\geq 0$ for all $k$, then the eigenvalues $\lambda$  are all positive real. 

One might wonder whether or not it is possible for $1<\ell<m-1$ to combine our method in this paper and the method in author's earlier paper \cite{Shin2}, in order to study the reality of the eigenvalues of the ($\mathcal{PT}$-symmetric) equation (\ref{shin_eq}) under the boundary condition (\ref{bdcond}). 
The work of Hille \cite[\S 7.4]{Hille} mentioned in Section \ref{properties} certainly covers the equation of the form (\ref{shin_eq}), and Sibuya \cite{Sibuya} studied  the more general equation  (\ref{shin_eq}) in a rotated form (even though in this paper we state only a special case, Proposition \ref{prop}, of his results for simplicity). So one can check that the material in Section \ref{properties} has a natural generalization (see Section 2 in \cite{Shin2} for some generalization). 

Moreover, the spectral determinants in these cases, corresponding to those in Lemmas \ref{equiv} and \ref{equiv1},  are entire functions of the coefficient vector $a:=(a_{m-1},a_{m-2},\cdots, a_1)\in \C^{m-1}$. Hence the eigenvalues $\lambda_j(a)$ are continuous in the coefficient vector $a\in \C^{m-1}$. With $a=0$ we know the $\lambda_j(0)$ are all positive real by Theorem \ref{main1} in this paper, and so for each $j\in\N$ there exists a  neighborhood of the origin in $\C^{m-1}$ such that if the coefficient vector $a$ lies in this  neighborhood, we have $|\arg \lambda_j(a)|\leq\frac{2\pi}{m+2}$. So if the coefficient vector $a$ lies in the intersection of all such  neighborhoods of the origin in $\C^{m-1}$, so that $|\arg \lambda_j(a)|\leq\frac{2\pi}{m+2}$ for all $j\in\N$, then 
using the ideas in this paper one would prove, with a little effort, the reality of the eigenvalues, provided $a\in\R^{m-1}$ with some sign restrictions on $a_k$ like for the $\ell=1$ case above. Here we restrict $a\in\R^{m-1}$ because this ensures that the eigenvalues appear in complex conjugate pairs or else are real, which is needed in the proof of the analogue of Theorem \ref{main}. This along with inequalities $|\arg \lambda_j(a)|\leq\frac{2\pi}{m+2}$ for all $j\in\N$ proves identities that correspond to $\left|D_{-n,n}(\omega^{-1}\lambda)\right|=1$
for $\ell=2n-1$  odd, and $\left|D_{-n-1,n}(\lambda)\right|=1$ for $\ell=2n$  even. Once we have these identities, the remaining portion of the proof of reality of the eigenvalues follows closely the method in \cite{Shin2}. However, {\it the author does not know whether or not the intersection of all such  neighborhoods of the origin in $\C^{m-1}$ contains more than just $a=0$.} 

If the coefficient vector $a\in\R^{m-1}$ lies outside the intersection mentioned in the previous paragraph, it seems that one should try to separately  use induction on $\ell$. 
Unlike for the case $P\equiv 0$, when $P\not\equiv 0$, if we try to establish an induction step similar to Theorem \ref{main}, the equation corresponding to (\ref{w_eq}) in the proof Theorem \ref{main} for $\ell=2n-1$ odd becomes
$$-w^\dd(z)+\left(z^m-\omega^{-1}P(\omega^{-\frac{1}{2}}z)+\omega^{-1}\lambda\right)w(z)=0,$$
with $w$ decaying in $S_{n+1}\cup S_{-n}$. So following the method in the proof of Theorem \ref{main}, we need to have $|\arg \sigma_j|\leq \frac{2\pi}{m+2}$ if $\sigma_j$ is an eigenvalue of  $H_{2n-1}^{P_{2n-1}}$ where  $P_{2n-1}(z):= -\omega^{-1}P(\omega^{-\frac{1}{2}}z)$ that is non-real.  There is no reason to believe the eigenvalues of $H_{2n-1}^{P_{2n-1}}$ (which is not $\mathcal{PT}$-symmetric) are all positive real. So there could perhaps be some non-real eigenvalues. A challange comes from not knowing a method of  estimating the arguments of the non-real eigenvalues.
The infinite product expressions like (\ref{prod_eq}) are very effective for proving the reality of the eigenvalues in this paper, but are they still useful in estimating the arguments of the non-real eigenvalues? Perhaps not.

In \cite{Shin}, the author studied the equation (\ref{shin_eq}) with $m$ odd, under the decaying boundary condition at the both ends of the real axis, and proved the eigenvalues $\lambda$ lie in the sector  $|\arg \lambda|\leq \frac{\pi}{m+2}$ under some restriction on the coefficient vector, but still with some large $a\in\R^{m-1}$. So we have a promising method of estimating the arguments of the non-real eigenvalues. But in the current case, we need to have a reverse induction on $1\leq\ell\leq\frac{m}{2}$. One can check with a little effort that even for $P\equiv 0$, this attempted reverse induction does not give us the desired identities $\left|D_{-n,n}(\omega^{-1}\lambda)\right|=1$
for $\ell=2n-1$ odd, and $\left|D_{-n-1,n}(\lambda)\right|=1$ for $\ell=2n$  even, that are needed in our proof of Theorem \ref{main}. So one  needs to show identities like these perhaps in a different way.

Finally, a simple way of seeing why this reverse induction on $1\leq\ell\leq\frac{m}{2}$ fails is as follows. In establishing the induction step in Theorem \ref{main}, we restrict $\ell$ on the interval $1\leq \ell\leq \frac{m}{2}-1$ because we have some technical difficulty that one can see from sentences right before equation (\ref{D_unit}), and right after  equation (\ref{D_unit21}).  From this observation, it is not difficult to see that trying to establish a reverse induction step on  $1\leq\ell\leq\frac{m}{2}$ is essentially the same as trying to establish the forwards induction step on $\frac{m}{2}<\ell\leq m-1$ that our method is unable to provide. (In proving Theorem \ref{main1} for $\frac{m}{2}<\ell\leq m-1$, we used the reflection $z\mapsto -z$.)

\section{Conclusions}
\label{conclusions}
In this paper, we proved that for each integer $1\leq \ell\leq m-1$, eigenvalues $\lambda\in\C$ of
$$-u^\dd(z)+(-1)^{\ell}(iz)^m u(z)=\lambda u(z),$$
under the boundary condition that $u(z)\rightarrow 0$ exponentially, as $z\rightarrow \infty$ along the two rays $\arg z=-\frac{\pi}{2}\pm\frac{\ell+1}{m+2}$, are all positive real. And we studied other boundary conditions. Due to asymptotic behavior of the solution $u(z)$ near infinity, if the Stokes sectors that contain the two ray where we impose the boundary condition are symmetric with respect to the imaginary axis, then the eigenvalues are all positive real, except the case when the two Stokes sectors contain the imaginary axis.  For all other boundary conditions, either there is no eigenvalue or the eigenvalues are not real.

It will be interesting to consider the eigenvalue problem with more general polynomial potentials $(-1)^{\ell} (iz)^m+P(iz)$ where $P(z)$ is a real polynomial with degree less than $m$, under the boundary condition (\ref{bdcond}). 

It is known that some $\mathcal{PT}$-symmetric oscillators with some cubic and quartic polynomial potentials have non-real eigenvalues \cite{Pham,Delabaere,Handy2,Handy1}. One would like to classify when  $\mathcal{PT}$-symmetric oscillators with  polynomial potentials have non-real eigenvalues. 

\subsection*{{\bf Acknowledgments}}

The author thanks Richard S. Laugesen for continued encouragement throughout the work.

{\sc email contact:}  kcshin@math.uiuc.edu
\end{document}